\documentclass[aps,prc,nofootinbib,amsmath,amssymb, onecolumn]{revtex4-2}
\usepackage{graphicx}% Include figure files
\usepackage{dcolumn}% Align table columns on decimal point
\usepackage{bm}
\usepackage{color}
\usepackage{mathrsfs}
\usepackage{comment}
\usepackage{mathtools}
\usepackage{eurosym}
\usepackage{amsmath,amssymb}
\usepackage{graphicx}
\usepackage{dcolumn}
\usepackage{bm}
\usepackage{color}
\usepackage{multirow}
\usepackage{threeparttable}

\usepackage{float}

\usepackage{array}
\usepackage{dcolumn}
\usepackage{booktabs}
\usepackage[detect-all]{siunitx}
\usepackage{makecell}
\usepackage{xcolor}
\usepackage{yfonts}
\newcolumntype{P}[1]{>{\centering\arraybackslash}p{#1}}
\begin{document}
\title{Magnetoexcitons in phosphorene monolayers, bilayers, and van der Waals heterostructures}
\author{Roman Ya. Kezerashvili$^{1,2}$ and Anastasia Spiridonova$^{1,2}\thanks{%
E-mail contact:}$  }
\affiliation{$^{1}$New York City College of Technology, The City University of New York, Brooklyn, NY 11201, USA\\
$^{2}$The Graduate School and University Center, The City University of New
York, New York, NY 10016, USA
}
\date{\today}
\begin{abstract}
We study direct and indirect excitons in Rydberg states in
phosphorene monolayers, bilayer and van der Waals (vdW) heterostructure in an external magnetic field, applied perpendicular to the
monolayer or heterostructure within the framework of the effective mass approximation. Binding energies of
magnetoexcitons are calculated by a numerical integration of the Schr\"{o}dinger equation
using the Rytova-Keldysh potential for direct magnetoexcitons and both the Rytova-Keldysh and Coulomb
potentials for indirect one. The latter allows to understand the role of screening in phosphorene. We report the magnetic field energy contribution to the binding energies and diamagnetic coefficients (DMCs) for magnetoexcitons that strongly depend on the effective mass of electron and hole and their anisotropy and can be tuned by the external magnetic field. We demonstrate theoretically that the vdW phosphorene heterostructure is a novel category of 2D
semiconductor offering a tunability of the binding energies of magnetoexcitons by mean of external magnetic field and 
control the binding energies and DMCs by the number of hBN layers separated two phosphorene sheets. Such tunability is potentially useful for the devices design.

\end{abstract}
\maketitle

\section{Introduction}

The most recent addition to the growing family of 2D material is phosphorene
that is the monolayer of black phosphorus (BP)\emph{. }The BP is composed of
individual phosphorene layers. The blossoming interest in phosphorene
derives partially from its direct gap that is retained in both monolayer and
bulk structure, high mobility, and high on-off current ratio in
field-effective transistors %
\citep{Qiao2014,Liu2014,Castellanos_Gomez_2014,Li2014}. In contrast, these
characteristics simultaneously are not present in graphine,
transitional metal dichalcogenides (TMDCs): WSe$_{2}$, WS$_{2}$, MoSe$_{2}$,
MoS$_{2}$, or Xenes: silicene, germanene, stanene, making phosphorene a
promising material for electronic and optical applications. Black phosphorus
is the most stable of phosphorus allotropes. It was first synthesized in
1914 by Bridgman \cite{Bridgman1914}. While bulk black phosphorus has
interesting properties and was extensively studied in twentieth century %
\citep{Keyes1953,Warschauer1953,	Cartz1979,Takao1981,Morita1986,Nolang1990},
it was not used in the design of electronic devices. 

However, everything has
changed after the discovery of graphene in 2004 \cite{Novoselov2004} which
paved the way for two-dimensional materials. Initially, the research was
focused on graphene \citep{Neto2009,Avouris2017} and TMDCs %
\citep{Cheng2014,Lee2014,Cotlet2016,Manzeli2017,Berman2016,Berman2017,Wang2018,Brunetti2018,RKAS2021b}
then silicene and other group VI elements have followed %
\citep{Matthes2013,Molle2017,Zheng2020,Kez2021}. There are some drawbacks in
the above materials. Graphene has high mobility and on-off ratio but a lack
of a gap impedes its use in the design of electronic devises \cite{Li2014, Liu2014}%
. TMDCs have a gap \citep{Mak2010,Splendiani2010,Tongay2012}, but its
carrier mobility is orders of magnitudes lower than in graphene. The
phosphorene that has been synthesized in 2014 has advantages over graphene,
TMDCs, and Xenes. Its most remarkable properties include thickness-dependent
band gap, strong in-plane anisotropy, and high carrier mobility.

BP has a structure where monolayer appears to be composed of two distinct
planes made of puckered honeycomb structure \cite%
{Carvalho2016,Batmunkh2016,Qiu2017}. This results in an anisotropic electronic
structure. Due to phosphorene unique topological structure and differences
between the armchair (AC) and zigzag (ZZ) directions, it displays strong
in-plane anisotropy. Many properties of phosphorene in these two principal
directions are drastically different. The anisotropic structure is strongly
reflected in effective masses of charge carriers and leads to anisotropic
effective masses of electrons and holes along AC and ZZ ($x$ and $y$)
directions. Along the AC direction the effective electron and hole masses are
smaller than along the ZZ direction
\citep{Takao1981,Carvalho2016,Batmunkh2016,
Qiu2017}. The distinct features of BP is that the direct gap located at the $%
\Gamma $ point is preserved in the monolayer and bulk structure. This
material is a direct band gap semiconductor with a
strongly anisotropic dispersion in the vicinity of the gap. It is worth
noting that the unusual structure of phosphorene sets it aside from graphene
and other widely studied 2D semiconductors.

Since the synthesis of phosphorene, it has been extensively studied. For
example, optical and thermal properties have been considered in Refs.\cite%
{Low2014,Ezawa2014,Qiu2017,Brunetti2019, Yoon2021}, in plane electric field
has been studied in Refs. \cite{Koenig2014,Chaves2015,Le2019,Kamban2020},
the effects of the strain on different properties of phosphorene have been
addressed in Refs. \citep{Rodin2014,Li2014b,Lv2014}, the Landau levels have
been reported in Refs. %
\citep{Zhou2015,Zhou2015b,Tahir2015,Faria2019,Zhou2020,Wu2020}.

Phosphorene hosts tightly bound excitons %
\citep{Tran2015,Wang2015,Carvalho2016,Tian2020,Henriques2020}. Similarly to
other monolayer semiconductors, reduced dimensionality and reduced screening
of the Coulomb attraction lead to a high exciton binding energy in
phosphorene\emph{. }Moreover,\emph{\ }the exciton binding energy in
phosphorene is larger than the one in other 2D materials. Excitons in
semiconductors in the presence of the external magnetic field have been
studied for the past sixty years. Elliot and Loudon \cite{Elliott1960} and
Hasegawa and Howard \cite{Hasegawa1961} developed the theory of the Mott
exciton in the strong magnetic field. Authors in Refs. \cite%
{Shinada1965,Gorkov1967, Akimoto1967} addressed Mott excitons. Excitons in
TMDCs monolayers, bilayer, and the double-layer heterostructure in the
presence of the external magnetic field have been extensively studied. %
The diamagnetic shifts have been reported in Refs. \cite%
{Walck,Luckert2010,
Choi2015,donckexc2018,DonckDM2018,Stier2018,Han2018,Spiridonova,RKAS2021b},
and Zeeman shifts have been reported in Refs \cite%
{Li2014zm,Srivastava2014,Aivazian2015,MacNeill2015,Plechinger2016, Stier2016}. We cite these works, but the recent literature on the subject is not limited by them. However, there is a lack of similar research on excitons in phosphorene.
This motivates us to study the effect of the external magnetic field on the
binding energies of Rydberg states of magnetoexcitons in monolayer,
bilayer, and the double-layer heterostructure composed of phosphorene and to
calculate diamagnetic coefficients (DMCs). 

In this paper, we study the dependence of the magnetoexciton binding energy of
Rydberg states: 1$s$, 2$s$, 3$s$, and 4$s$, on the external magnetic field perpendicular to a monolayer or heterostructure
that is varied between 0 and 60 T and report the diamagnetic coefficients
(i) for the direct magnetoexcitons in a freestanding (FS) and encapsulated phosphorene monolayers, (ii) for the indirect
magnetoexciton in bilayer composed of two phosphorene monolayers, (iii)
for the indirect magnetoexcitons in heterostructure formed by two
phosphorene monolayers and separated by $N$ hBN monolayers. The latter van der Waals 
heterostructure is denoted as vdW. 
%and the number of hBN layers is varied between 1 and 6. 
In our approach, we numerically solve the Schr\"{o}%
dinger equation for the magnetoexciton and obtain eigenfunctions and
eigenvalues. Then we obtain the energy contribution from the magnetic field
to the binding energies and use it to calculate diamagnetic coefficients.
For the direct exciton, we solve the Schr\"{o}dinger equation with the
Rytova-Keldysh potential \cite{Rytova,Keldysh}, and for the indirect
exciton, we solve the Schr\"{o}dinger equation with the Rytova-Keldysh and
the Coulomb potentials to understand the role of the screening in phosphorene.

The remainder of this paper is organized in the following way. In section \ref{theory}
the theoretical model for the description of an electron-hole system in the
external magnetic field with the charge carriers effective mass anisotropy
is presented. Results of calculations of binding energies of Rydberg states and DMCs for direct magnetoexcitons in a freestanding and encapsulated
phosphorene and indirect magnetoexcitons in FS bilayer and vdW phosphorene 
heterostructures are presented in Sec. \ref{results}. Finally, conclusions follow in  Sec. \ref{coefficients}.

\section{Theoretical Model}

\label{theory}

It is known that electrostatically-bound electrons and holes in the external
magnetic field form magnetoexcitons. In this section, following Refs. \cite%
{Gorkov1967,Spiridonova,Kez2021,RKAS2021b}, we introduce briefly the
theoretical model for the description of the Mott-Wannier magnetoexciton in
phosphorene. We consider the energy contribution from the external magnetic
field to the Rydberg states binding energies of magnetoexcitons and
diamagnetic coefficients (DMCs). In the system under consideration, excitons are confined in a 2D freestanding and encapsulated
phosphorene monolayer, FS bilayer phosphorene and van der Waals
heterostructure, where $N$ layers of hBN monolayers separate two phosphorene
monolayers. In the latter two cases equal number of electrons and holes are
located in parallel phosphorene monolayers at a distance $D$ away. The
corresponding schematic illustrations of these systems are shown in Fig. \ref%
{fig1M}.

It is worth mentioning that we are considering the monolayers of hBN as an
insulator. Phosphorene encapsulated between hBN layers is robust to
oxidation and exhibits high mobilities. Moreover, hBN also has a high
dielectric constant, resulting in a strong damping of the electrostatic
repulsion by charged impurities which are responsible for the decrease of
carriers mobility.
\begin{figure}[b]
\begin{centering}
\includegraphics[width=14.0cm]{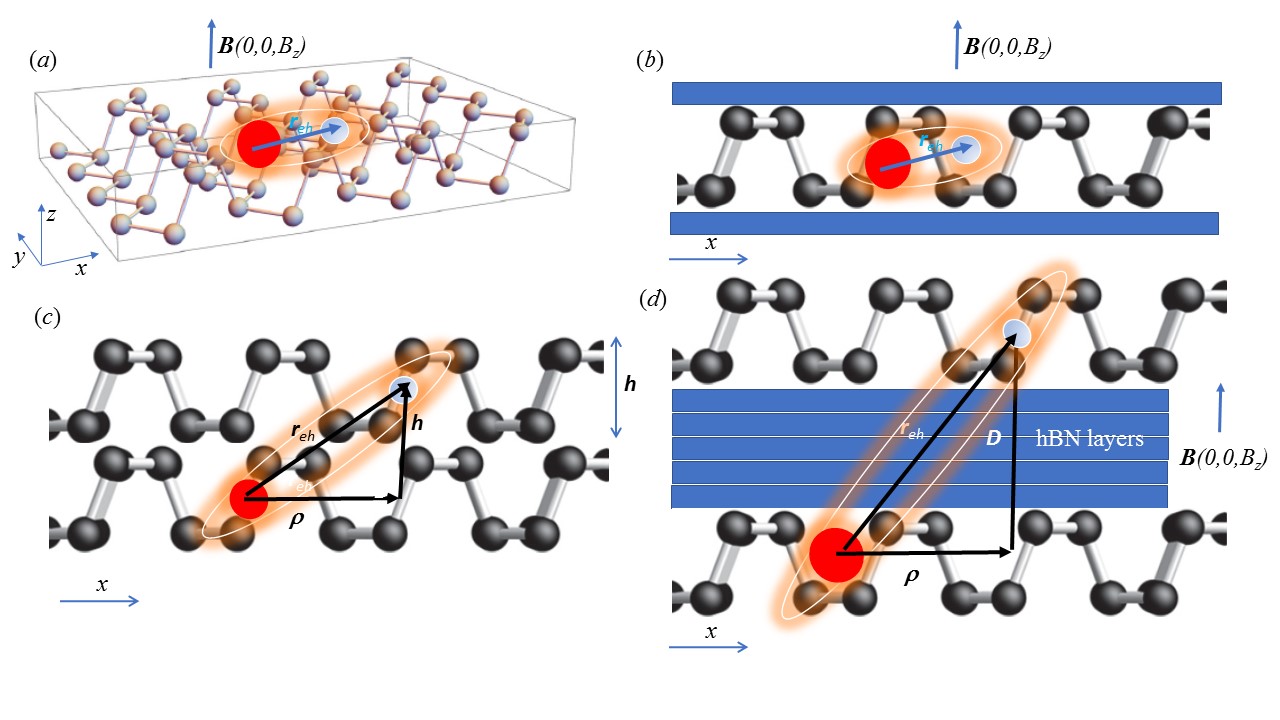}
\caption{(Color online) Schematic illustration of magnetoexcitons in phosphorene monolayers and heterostructures. $(a)$ A direct magnetoexciton in a freestanding phosphorene monolayer. $(b)$ A direct magnetoexciton in an encapsulated phosphorene monolayer. $(c)$ An indirect magnetoexciton in a freestanding bilayer phosphorene heterostructure. $(d)$ An indirect magnetoexciton in phosphorene van der Waals heterostructure.}
\label{fig1M}
\end{centering}
\end{figure}

Let us introduce the coordinate vectors of the electron and hole for the
Mott-Wannier exciton in the phosphorene layer. The following in-plane
coordinates $\mathbf{r}_{1}=(x_{1},y_{1})$ and $\mathbf{r}_{2}=(x_{2},y_{2})$
for an electron and hole, respectively, will be used in our description.
We assume that at low momentum $\mathbf{p}=(p_{x},p_{y})$, i.e., near the $%
\Gamma$ point, the single electron and hole energy spectrum $\varepsilon
_{l}^{(0)}(\mathbf{p})$ is given by

\begin{eqnarray}
\varepsilon_{l}^{(0)}(\mathbf{p}) = \frac{p_{x}^{2}}{2m_{x}^{l}} + \frac{
p_{y}^{2}}{2m_{y}^{l}}, \ \ l =e,\ h,  \label{esingl}
\end{eqnarray}
where $m_{x}^{l}$ and $m_{y}^{l}$ are the electron and hole effective masses
along the $x$ and $y$ directions, respectively. We assume that $OX$ and $OY$
axes correspond to the armchair and zigzag directions in a phosphorene
monolayer, respectively.

The anisotropic nature of the 2D phosphorene atomic semiconductor, in
contrary to other 2D isotropic materials such as graphene and TMDC semiconductors,
breaks the central symmetry and requires for the description of excitons the use of
the Cartesian coordinates. The asymmetry of the electron and hole dispersion
in phosphorene is reflected in the Hamiltonian for the Mott-Wannier
magnetoexciton, and within the framework of the effective mass approximation
the Hamiltonian %Schr\"{o}dinger equation
for an interacting electron-hole pair in phosphorene in the external
magnetic field reads ($\hbar=c=1$)
\begin{eqnarray}
\hat{H}=&&\frac{1}{2m_{e}^{x}}\Big( i\nabla _{x_{e}}-eA_{x}(r_{e})\Big) ^{2}+%
\frac{1}{2m_{e}^{y}}\Big( i\nabla _{y_{e}}-eA_{y}(r_{e})\Big) ^{2}+\frac{1}{%
2m_{h}^{x}}\Big( i\nabla _{x_{h}}+eA_{x}(r_{h})\Big) ^{2}+\frac{1}{%
2m_{h}^{y}}\Big(  i\nabla _{y_{h}}+eA_{y}(r_{h})\Big) ^{2}  \notag \\
&&+V\Big( |\mathbf{r}_{e}-\mathbf{r}_{h}|\Big),  \label{eq:Schredingermag}
\end{eqnarray}
where the $m_{i}^{j},~j=x,y,~i=e,h$ correspond to the effective mass of the
electron or hole in the $x$ or $y$ direction, respectively, and $%
V\left( |\mathbf{r}_{e}-\mathbf{r_{h}}|\right) $ describes the electrostatic
interaction between the electron and hole. When the electron and hole are
located in 2D plane, we use the Rytova-Keldysh (RK) potential \cite%
{Rytova,Keldysh} that is widely used for the description of charge carriers
interaction in 2D materials. The RK potential is a central potential, and the
interaction between the electron and hole for direct excitons in a
monolayer has the form \cite{Rytova, Keldysh}:
\begin{equation}
V_{RK}(r)=-\frac{\pi ke^{2}}{2\kappa \rho _{0}}\left[ H_{0}\left( \frac{r}{%
\rho _{0}}\right) -Y_{0}\left( \frac{r}{\rho _{0}}\right) \right] ,
\label{eq:rk}
\end{equation}%
where $r=r_{e}-r_{h}$ is the relative coordinate between the electron and
hole. In Eq.~(\ref{eq:rk}), $e$ is the charge of the electron, $\kappa
=(\epsilon _{1}+\epsilon _{2})/2$ describes the surrounding dielectric
environment, $\epsilon _{1}$ and $\epsilon _{2}$ are the dielectric
constants below and above the monolayer, $H_{0}$ and $Y_{0}$ are the Struve and Bessel
functions of the second kind, respectively, and $\rho _{0}$ is the screening
length. The screening length $\rho _{0}$ can be written as $\rho _{0}=2\pi \chi _{2D}/\kappa $ ~\cite%
{Berkelbach2013}, where $\chi _{2D}$
is the 2D polarizability, which in turn is given in \cite{Keldysh}. At the long-range distances $V_{RK}$ retains  $1/r$ behavior as the Coulomb potential, for smaller distances the potential well is logarithmic. 

%and $\Psi (\mathbf{r}_{1},\mathbf{r}_{2})$ and $E$ are the eigenfunction and eigenenergy.
Following the standard procedure~\cite{Landau} for the separation of the
relative motion of the electron-hole pair from their center-of-mass motion
one introduces variables for the center-of-mass of an electron-hole pair
$\mathbf{R}=(X,Y)$ and the relative motion of an electron and a hole $%
\mathbf{r}=(x,y)$, as $%
X=(m_{x}^{e}x_{1}+m_{x}^{h}x_{2})/(m_{x}^{e}+m_{x}^{h})$, $%
Y=(m_{x}^{e}y_{1}+m_{x}^{h}y_{2})/(m_{x}^{e}+m_{x}^{h})$, $x=x_{1}-x_{2}$, $y=y_{1}-y_{2}$, $r^{2}=x^{2}+y^{2}$. The Schr\"{o}dinger equation with
Hamiltonian~(\ref{eq:Schredingermag}) has the form: $\hat{H}\Psi (\mathbf{r}%
_{1},\mathbf{r}_{2})=\mathcal{E}\Psi (\mathbf{r}_{1},\mathbf{r}_{2})$, where
$\Psi (\mathbf{r}_{1},\mathbf{r}_{2})$ and $\mathcal{E}$ are the
eigenfunction and eigenenergy, respectively. One can write $\Psi (\mathbf{r}_{1},\mathbf{r}%
_{2})$ in the form $\Psi (\mathbf{r}_{1},\mathbf{r}_{2})=\Psi (\mathbf{R},%
\mathbf{r})=e^{i\mathbf{P}\cdot \mathbf{R}/\hbar }\Phi (x,y)$, where $%
\mathbf{P}=(P_{x},P_{y})$ is the momentum of the center-of-mass of the
electron-hole pair and $\Phi (x,y)$ is the wave function of the relative motion of the
electron-hole pair. Following Refs. \cite%
{MacDonald1986,Spiridonova,Kez2021,RKAS2021b}, after lengthy
calculations one obtains the equation that describes the Mott--Wannier
exciton in Rydberg optical states in the external magnetic field
perpendicular to the phosphorene monolayer. Finally, the equation for the
relative motion of the electron and hole in the phosphorene monolayer with
zero center-of-mass momentum reads
\begin{equation}
\left[ -\frac{1}{2 \mu_x}\frac{\partial ^2}{\partial x^2}-\frac{1}{2\mu_y}%
\frac{\partial ^2}{\partial y^2}+\frac{e^2}{8 \mu_x}B^2x^2+\frac{e^2}{8 \mu_y%
}B^2 y^2+V(x,y)\right] \Phi (x,y)=E\Phi (x,y).  \label{eq:finsch}
\end{equation}%
where $\mu_{x} = \frac{m_{x}^{e} m_{x}^{h}}{m_{x}^{e}+ m_{x}^{h}}$ and $%
\mu_{y} = \frac{m_{y}^{e} m_{y}^{h}}{m_{y}^{e}+ m_{y}^{h}}$ are the reduced
masses, related to the relative motion of an electron-hole pair in the $x$
and $y$ directions, respectively. In Eq. (\ref{eq:finsch}) the anisotropy is present in the kinetic and magnetic terms, while the
potential term has isotropic form.

\begin{figure}[t]
\begin{centering}
\includegraphics[width=9.0cm]{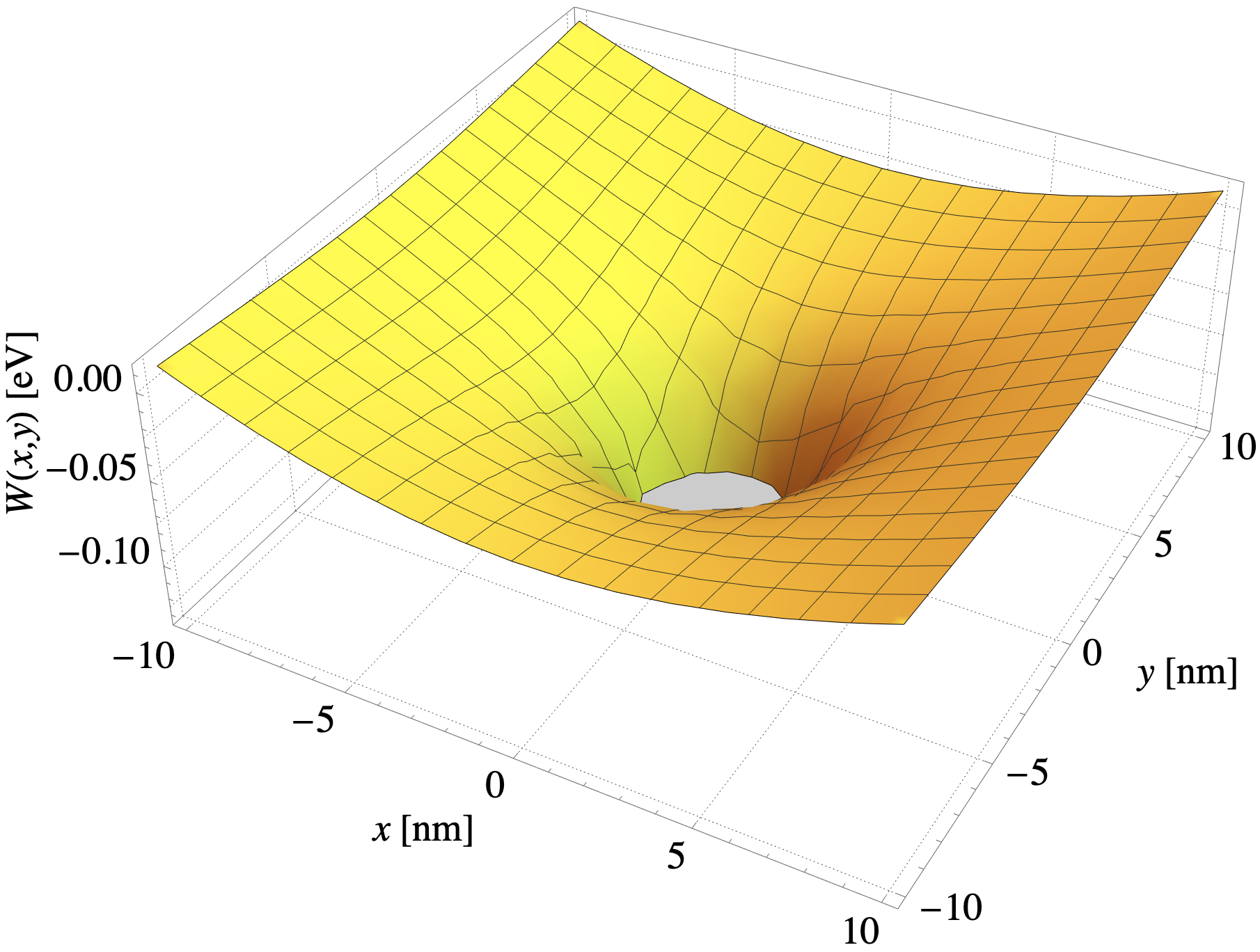}
\caption{Total potential $W(x,y)=V_{RK}(x,y)+\frac{e^2 }{8 \mu_x}B^2 x^2+\frac{e^2}{8 \mu_y}  B^2y^2$ is plotted as a function of $x$ and $y$. $W(x,y)$ and $x$ and $y$ are given in eV and nm, respectively. The potential is calculated at $B=30$ T with the parameters of the reduced masses of the set 1.}
\label{potential}
\end{centering}
\end{figure}

Following Refs. \cite%
{Kezerashvili2019,RK2020} one can obtain from (\ref{eq:finsch}) the
expectation value for the ground state energy. For the case of $V_{RK}(x,y)$ potential we have $E=\left\langle -\frac{1}{%
2\mu _{x}}\frac{\partial ^{2}}{\partial x^{2}}\right\rangle +\left\langle -%
\frac{1}{2\mu _{y}}\frac{\partial ^{2}}{\partial y^{2}}\right\rangle
+\left\langle \frac{e^{2}}{8\mu _{x}}B^{2}x^{2}\right\rangle +\left\langle
\frac{e^{2}}{8\mu _{y}}B^{2}y^{2}\right\rangle +\left\langle
V_{RK}(x,y)\right\rangle $. The later expression could be viewed as the sum of
the average values of the operators of kinetic and potential energies in 2D
space obtained by using the corresponding eigenfunction $\Phi (x,y)$ of the
exciton. Phosphorene exhibits charge carriers effective mass anisotropy with
lighter effective masses along the armchair direction and heavier effective
masses along the zigzag direction that leads to the reduced mass anisotropy:%
\emph{\ }$\mu _{y}>\mu _{x}$. For the set of masses presented in Table \ref%
{table:param} the ratio $\mu _{y}/\mu _{x}$ varies from $\sim $ 7 to 16.
Therefore, from the latter expression one can conclude that contributions
from the kinetic and magnetic energies to the total ground state energy can
be by the order of magnitude different for AC and ZZ directions. In particular, the kinetic energy along the AC direction is much larger than that along the ZZ direction. Thus, the
anisotropic nature of the 2D phosphorene atomic semiconductors, in contrast
to other 2D materials such as graphene, Xenes, and TMDC semiconductors,
allows excitons to be confined in a quasi-one-dimensional space predicted in
theory \cite{Tran2014, Qiao2014}, leading to remarkable phenomena arising
from the reduced dimensionality and screening. In Fig. \ref{potential} the total potential $W(x,y)=V_{RK}(x,y)+\frac{e^2 }{8 \mu_x}B^2 x^2+\frac{e^2}{8 \mu_y} B^2 y^2$ that acts on the electron-hole system in phosphorene is presented. One can observe the anisotropic structure of this potential and its asymmetry with respect to the AC and ZZ directions.

Let us now to consider the indirect magnetoexcitons formed by electrons and
holes located in two different phosphorene monolayers in the bilayer or vdW heterostructure. In the latter case the phosphorene layers are
separated by $N$ layers of hBN monolayers. Such magnetoexcitons have a
longer lifetime than the direct excitons due to longer recombination time.
Equation (\ref{eq:finsch}) still describes the indirect exciton. However,
for indirect excitons, the expressions for the interaction between the
electron and hole in Eq. (\ref{eq:finsch}) can be written as:
\begin{equation}
V_{RK}(\sqrt{\rho ^{2}+D^{2}})=-\frac{\pi ke^{2}}{2\kappa \rho _{0}}\left[
H_{0}\left( \frac{\sqrt{\rho ^{2}+D^{2}}}{\rho _{0}}\right) -Y_{0}\left(
\frac{\sqrt{\rho ^{2}+D^{2}}}{\rho _{0}}\right) \right] ,  \label{eq:indkeld}
\end{equation}%
for the RK potential, and
\begin{equation}
V_C \left( \sqrt{\rho ^{2}+D^{2}}\right) =-\frac{ke^{2}}{\kappa \left( \sqrt{%
\rho ^{2}+D^{2}}\right) }  \label{eq:indcoul}
\end{equation}%
for the Coulomb potential and where $\rho ^2 = x^2+y^2$. Equations~\eqref{eq:indkeld} and~%
\eqref{eq:indcoul} describe the interaction between the electron and hole
that are located in different parallel phosphorene monolayers separated by a
distance $D=h+Nl_{\text{hBN}}$, where $l_{\text{hBN}}=0.333$ nm is the thickness of the hBN layer and $h$ is the phosphorene thickness given in Table \ref{table:param}. Therefore, one can obtain the eigenfunctions and eigenenergies
of magnetoexcitons by solving Eq.~(\ref{eq:finsch}) using the potential~(%
\ref{eq:rk}) for direct magnetoexcitons, or using either potential~(\ref%
{eq:indkeld}) or~(\ref{eq:indcoul}) for indirect magnetoexcitons.

\begin{table}[h]
\caption{Parameters for phosphorene. Four sets of reduced masses for phosphorene that are used in calculations as input parameters. The
anisotropic reduced masses $\protect\mu_x$ and $\protect\mu_y$ are obtained
based on effective masses of the electron and hole given in Ref.
\protect\cite{Peng2014} for the set 1, in Ref. \protect\cite{Tran2014-2} for the set
2, in Ref. \protect\cite{Paez2014} for the set 3, and in Ref.\protect\cite%
{Qiao2014} for the set 4. $\protect\mu_{x}$ and $\protect\mu_{y}$ are in units of the electron mass, $m_0$. $\chi_{2D}$ and $h$ are the polarizability and thickness of phosphorene, respectively, in nm.}
\label{table:param}
\centering
 \begin{threeparttable}[b]
\sisetup{table-format=2.4}
\begin{tabular}{P{1.5cm}P{1.5cm}P{1.5cm}P{1.5cm}P{1.5cm}P{1.5cm}}
\hline \hline
  & $\mu_x$ & $\mu_y$ & $\mu_y$/$\mu_x$& $\chi_{2D}$ (nm) & $h$ (nm)\\ \hline
 set 1 & 0.06296 & 0.96774 & 15.37 & \multirow{4}{*}{0.41\tnote{a}} & \multirow{4}{*}{0.541\tnote{b} } \\
 set 2 & 0.06667 & 0.88780 & 13.32 \\
 set 3 & 0.09122 & 0.6599 & 7.23  \\
 set 4 & 0.07969 & 0.9498 &  11.92 \\  \bottomrule \bottomrule
\end{tabular}

		\begin{tablenotes} \footnotesize
            \item [a] Reference \cite{Rodin2014b}.
            \item [b] Reference \cite{Kumar2016}.
        \end{tablenotes}
\end{threeparttable}
\
\end{table}

\section{Results of calculations and discussion} \label{results}

We report the dependence of the energy contribution from the external magnetic field to the
binding energies of magnetoexcitons in Rydberg states 1$s$, 2$s$, 3$s$, and 4%
$s$ in the FS (Fig. \ref{fig1M}($a$)) and encapsulated by hBN (Fig. \ref{fig1M}($b$)) phosphorene, the FS phosphorene bilayer (Fig. \ref{fig1M}($c$)), and
the vdW heterostructure (Fig. \ref{fig1M}($d$)) on the magnetic field. The diamagnetic coefficients for the monolayers, bilayer, and vdW phosphorene heterostructure are reported for the first time.

In calculations, we use effective masses for electron and hole found in
literature Refs.~\cite{Peng2014,Tran2014-2,Paez2014,Qiao2014}, which
are obtained by using the first principle calculations. The lattice
constants in Refs.~\cite{Peng2014,Tran2014-2,Paez2014,Qiao2014} do not
coincide with each other and different functionals for the correlation
energy and setting parameters for the hopping lead to some difference in
results for anisotropic masses. It is obvious that these can cause the
difference in the band curvatures and, therefore, effective masses along
the armchair and the zigzag directions. The latter motivate us to use in
calculations of binding energies of magnetoexcitons and DMCs the different
sets of masses from Refs.~\cite{Peng2014,Tran2014-2,Paez2014,Qiao2014} that
allows to understand the dependence of the binding energy of magnetoexcitons
and DMCs on effective masses of electrons and holes and the role of
anisotropic masses. The corresponding reduced masses $\mu _{x}$ and $\mu _{y}$ along
the armchair and zigzag directions, respectively, are given in Table \ref{table:param}.
We use these masses as input parameters and refer to them as the set 1, the set 2, the set
3, and the set 4. Below, we report results for the sets of masses 1 and 3
since they give an upper and lower bounds on the binding energies and the
energy contribution from the magnetic field to the binding energies of the
Rydberg states. Results from the sets of masses 2 and 4 fall within results of the
sets 1 and 3.

A numerical solution of the Schr\"{o}dinger equation (\ref{eq:finsch}), using
the aforementioned interaction potentials and input parameters from Table \ref{table:param},  
is performed using the finite element method, which yields pairs of
eigenenergies and eigenfunctions which are solutions to the Schr\"{o}dinger
equation, corresponding to the most-strongly-bound states. The method is
based on using the finite element method implemented in Wolfram Mathematica
in the NDEigensystem function. We modify the code successfully implemented
in Ref. \cite{Brunetti2019} in a way that it explicitly contains $\frac{e^{2}}{8\mu _{x}}%
B^{2}x^{2}$ and $\frac{e^{2}}{8\mu _{y}}B^{2}y^{2}$ terms. Our results for
the binding energies of direct excitons in FS and encapsulated hBN
phosphorene monolayers and indirect excitons in FS phosphrene bilayer
for the masses sets 1 and 3 are given in Table \ref{table:energies_mon_bil}. One can see that the binding energy of excitons in FS and encapsulated monolayers significantly depend on the anisotropic reduced masses. Moreover, a much lower screening of excitons in the freestanding monolayer makes their binding energies, for example, in the ground state over three fold bigger than when the monolayer is encapsulated by hBN. 

To check the validity of the code, we use the input parameters from respective papers listed
below and calculate the binding energies of direct and indirect excitons.
The code reproduces the theoretical binding energies of direct excitons in
phosporene and of indirect excitons in bilayer reported in Ref. \cite%
{Castellanos_Gomez_2014}, where binding energies are calculated using
Wannier effective-mass theory in 2D space, within 8\%. The binding energies for 1$s
$ state obtained using a semi-analytical perturbation theory approach \cite%
{Henriques2020} are reproduced within 4\%. The code reproduces binding
energies for $n=0,...,4$ states for FS phosphorene, phosphorene on SiO$_{2}$ substrate, and encapsulated by hBN phosphorene when energies obtained using numerical
procedure where the Schr\"{o}dinger equation is solved via a finite element
representation of the exciton wave equation \cite{Kamban2020_phos} within
1\% and using $k\cdot p$ theory \cite{Junior2019} within 4\%.

\begin{table}[tbp]
\caption{Binding energies of direct magnetoexcitons in FS and encapsulated
hBN phosphorene monolayers and indirect magnetoexcitons in FS phosphorene
bilayer for the sets of masses 1 and 3. For indirect magnetoexcitons in the
bilayer phosphorene binding energies are calculated using $V_{RK}$ and $V_C$ potentials
. Energies are given in meV.}
\label{table:energies_mon_bil}
\begin{center}
\sisetup{table-format=2.4}
\par
\begin{tabular}{P{1cm}P{1.5cm}P{1.5cm}P{1.5cm}P{1.5cm}|P{1.5cm}P{1.5cm}P{1.5cm}P{1.5cm}}
\hline \hline
\multirow{3}{*}{State}&
 \multicolumn{4}{c|}{Monolayer}&
  \multicolumn{4}{c}{Bilayer} \\ \cline{2-9}

& \multicolumn{2}{c}{Set 1}&
\multicolumn{2}{c|}{Set 3} &
\multicolumn{2}{c}{Set 1} &
\multicolumn{2}{c}{Set 3}
   \\ \cline{2-9}
  &  hBN & FS & hBN & FS & $V_{RK}$ & $V_C$ & $V_{RK}$ & $V_C$ \\ \cline{1-9}
1$s$ & 187.94 & 718.71 & 199.66 & 746.06 & 388.88 & 827.17 & 395.54 & 854.03\\
2$s$ & 78.62  & 488.84 & 74.37  & 478.75 & 298.56 & 653.05 & 294.50 & 644.97\\
3$s$ & 53.83  & 394.04 & 50.35  & 377.72 & 248.52 & 525.71 & 239.53 & 501.07\\
4$s$ & 35.67  & 319.55 & 32.03  & 311.22 & 211.61 & 430.68 & 209.07 & 429.89
\\   \hline \hline

\end{tabular}
\end{center}
\end{table}

\subsection{Contribution from the external magnetic field to binding
energies of magnetoexcitons in a monolayer}
\begin{figure}[b]
\begin{centering}
\includegraphics[width=16.0cm]{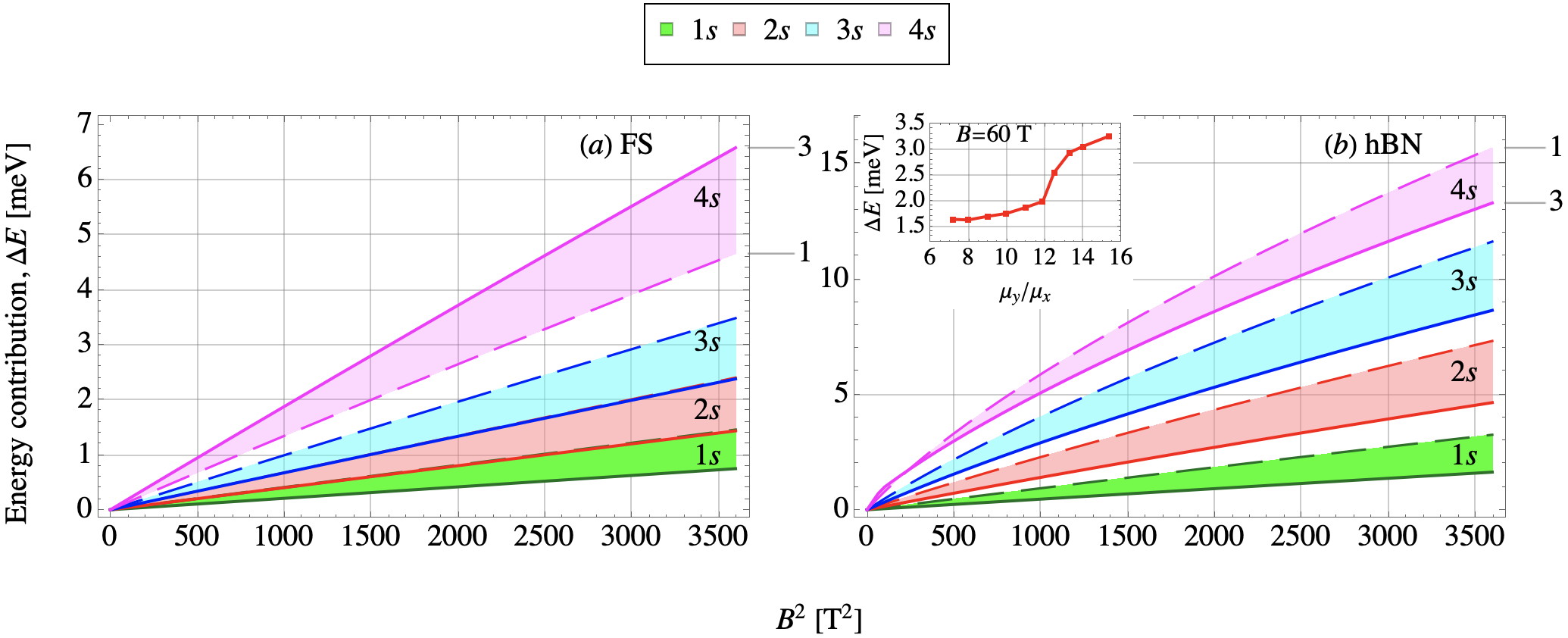}
\caption{Dependencies of the energy contribution from the magnetic field to the binding energies of magnetoexcitons in states 1$s$, 2$s$, 3$s$, and 4$s$ for FS ($a$) and encapsulated by hBN ($b$) phosphorene monolayers on the squared magnetic field. The boundary dashed and solid curves correspond to the sets of masses 1 and set 3, respectively. The contributions for the sets of masses 2 and 4 fall within the shaded region. The inset in Fig. \ref{monolayer_hbn_fs}($b$) shows the dependence of the energy contribution from the magnetic field on the ratio $\frac{\mu_{y}}{\mu_{x}}$.}
\label{monolayer_hbn_fs}
\end{centering}
\end{figure}
In Fig. \ref{monolayer_hbn_fs} ($a$)-($b$) we report the energy contribution
from the external magnetic field to the binding energies of Rydberg states
for direct magnetoexcitons in FS and encapsulated by hBN
phosphorene, respectively, as a function of $B^2$. We report results for
the sets of masses 1 and 3. The dashed and solid boundary curves in Fig. \ref{monolayer_hbn_fs} correspond
to the results obtained with the sets of masses 1 and 3, respectively. The results from the sets of masses 2 and 4 fall in the shaded region.

Interestingly enough, in phosphorene, we observe the same tendencies for magnetoexcitons as in
TMDCs \cite{Spiridonova,RKAS2021b} and Xenes \cite{Kez2021} monolayers. The binding energies and the energy contribution
from the magnetic field to the binding energies of Rydberg states of
magnetoexcitons strongly depend on the exciton reduced mass, even
though phosphorene has anisotropic mass along $x$ and $y$ directions. The reduced mass of the exciton along the AC direction, depending on the set, is
between seven and fifteen times smaller than the reduced mass along the ZZ
direction. Since according to Eq. (\ref{eq:finsch}), the Schr\"{o}dinger
equation contains terms that are proportional to $\frac{1}{\mu_{x}}$ and $\frac{1}{\mu_{y}}$, the terms with
significantly smaller $\mu_x$ dominate in Eq. (\ref{eq:finsch}). This is consistent with the anisotropic behavior of the potential $W(x,y)$
shown in Fig. \ref{potential} that exhibits the anisotropy with respect to the armchair and zigzag directions. Since the set
1 has the lowest $\mu_x$ and the highest $\mu_y$, the set 3 is vice verse, we
report results for the sets 1 and 3 as representative cases.

Here should be noted that in contrast to binding energies of direct excitons
in TMDCs and Xenes, the binding energies of direct excitons in phosphorene
have more complicated dependence on the anisotropic reduced masses. In phosphorene the central symmetry is strongly broken due to the large anisotropy of the electron and hole effective masses. This can
observed, for example, in the encapsulated by hBN phosphorene. The insert in Fig. \ref{monolayer_hbn_fs}$(b)$ shows the dependence of the magnetic field contribution to the binding energy of the direct magnetoexciton on the ratio $\frac{\mu_{y}}{\mu_{x}}$. Here we plot $\Delta E$ using the values $\frac{\mu_{y}}{\mu_{x}}$ from Table \ref{table:param}. $\Delta E$ for $\frac{\mu_{y}}{\mu_{x}} = 8,9,10,11,12.5$ and 14 are calculated by extrapolation of $\mu_{x}$ from  Refs.~\cite{Peng2014,Tran2014-2,Paez2014,Qiao2014} and obtaining corresponding $\mu_{y}$ using the value of $\frac{\mu_{y}}{\mu_{x}}$. The bigger $\frac{\mu_{y}}{\mu_{x}}$ ratio corresponds to the more anisotropic system that becomes quasi-one-dimensional \cite{Peng2014}, and the increase of $\frac{\mu_{y}}{\mu_{x}}$ leads to the increase of the contribution due to the magnetic field to the binding energy of the magnetoexciton. The character of these magnetoexcitons is that their spatial distribution of wave functions is strongly anisotropic. These magnetoexcitons form striped-like patterns, similar to those in nanotubes \cite{Bondarev2011} or nanowires \cite{RKezzaal2019} that are 1D systems.

Interestingly, the set 1 with the lowest $\mu_x$ and the highest $\mu_y$ has the lowest
1$s$ state binding energy, but for states 2$s$, 3$s$, and 4$s$ the set 3 with
the highest $\mu_x$ and the lowest $\mu_y$ gives the lowest binding energies.
The binding energy percent difference between the sets 1 and 3 in states 2$s$, 3$s$, and 4$s$
is 6\%, 7\%, and 11\%, respectively. Also, based on Fig. \ref%
{monolayer_hbn_fs}($a$), for states 1$s$-3$s$ the set 1 gives the highest energy
contribution to the binding energy and the set 3 gives the lowest $\Delta E$. However, for
the state 4$s$ the trend is reversed, and the set 3 gives the highest $\Delta E$, and
the set 1 gives the lowest $\Delta E$. In addition, in contrast to magnetoexcitons that
dissociate in the TMDCs and Xenes monolayers encapsulated by hBN \citep{Kez2021}, in states 3%
$s$ and 4$s$ the magnetoexcitons in phosphorene monolayers stay bound
when the magnetic field is varied in the range from 0 to 60 T. This feature is also the result of quasi-1D character of magnetoexcitons in phosphorene.

\subsection{Contribution from the external magnetic field to the binding
energies of magnetoexcitons in bilayer and vdW heterostructure}

\begin{figure}[h! ]
\begin{centering}
\includegraphics[width=16.0cm]{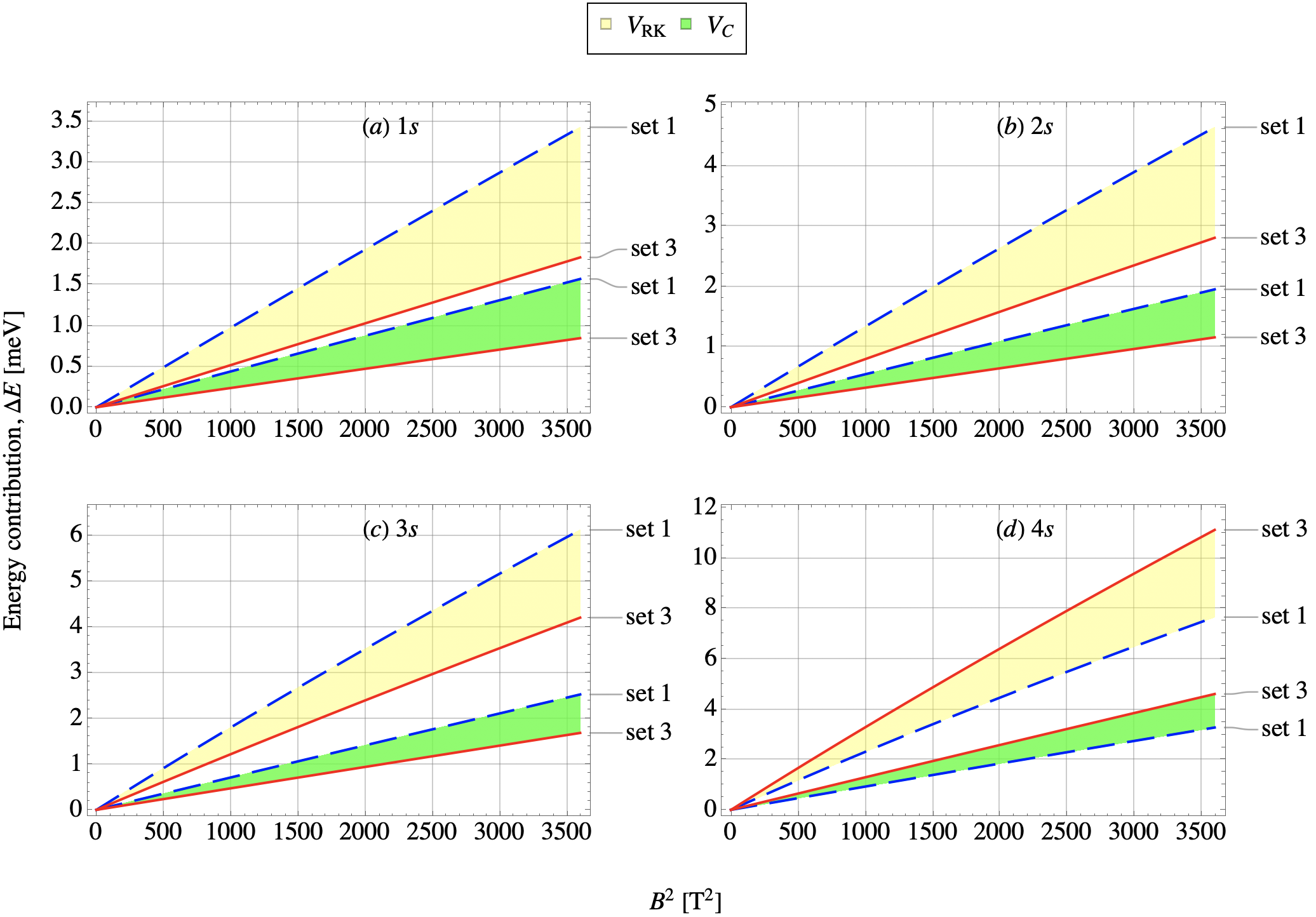}
\caption{The energy contribution from the magnetic field to the binding energy of Rydberg states 1$s$-4$s$ for the FS bilayer obtained using $V_{RK}$ and $V_C$ potentials. The boundary dashed and solid curves correspond to the sets of masses 1 and 3, respectively. The yellow and green shaded areas correspond to the calculations with $V_{RK}$ and $V_C$ potentials, respectively.}
\label{bilayer_min_max}
\end{centering}
\end{figure}
For indirect magnetoexcitons in the FS bilayer phosphorene and the vdW
heterostructure, we examine how interaction potential affects Rydberg states
binding energies and the energy contribution from the external magnetic
field to the binding energies. In addition, for the vdW
heterostructure, we consider an additional degree of freedom to tune binding
energies by number of hBN layers that separate two phosphorene monolayer.

In Fig. \ref{bilayer_min_max}, we report results for FS bilayer for the sets 1
and 3. The set 1 gives an upper bound on the energy contribution and is denoted
by the dashed line. The set 3 gives a lower bound on $\Delta E$ and is denoted
by the solid line. As was observed in the monolayer, in the state 4$s$ in
bilayer there is a flip of boundaries: the set 1 gives a lower boundary and the set
3 gives an upper boundary. Similar to results reported for magnetoexcitons
in Xenes and TMDCs bilayer systems in Ref. \citep{Kez2021,RKAS2021b}%
, in phosphorene bilayer, the following relation always holds $\Delta E _{RK}> \Delta E_C$. In addition, it is worth mentioning that
energy contribution from the magnetic field to the binding energy is higher
for indirect magnetoexcitons in the FS bilayer system than for direct
magnetoexcitons in FS phosphorene.
Based on the results in Table \ref%
{table:energies_mon_bil} for the bilayer phosphorene, Rydberg states binding energies calculated with
the Coulomb potential are twice as big as the binding energies calculated
with the Rytoval-Keldysh potential. Similar to the monolayer system,
magnetoexcitons in the bilayer in states 2$s$, 3$s$, and 4$s$ are more bound
than magnetoexcitons in TMDCs bilayer system.

\begin{figure}[b]
\begin{tabular}{cc}
\textit{(a)} 1$s$ & \textit{(b)} 2$s$ \\
\ \includegraphics[width=80mm]{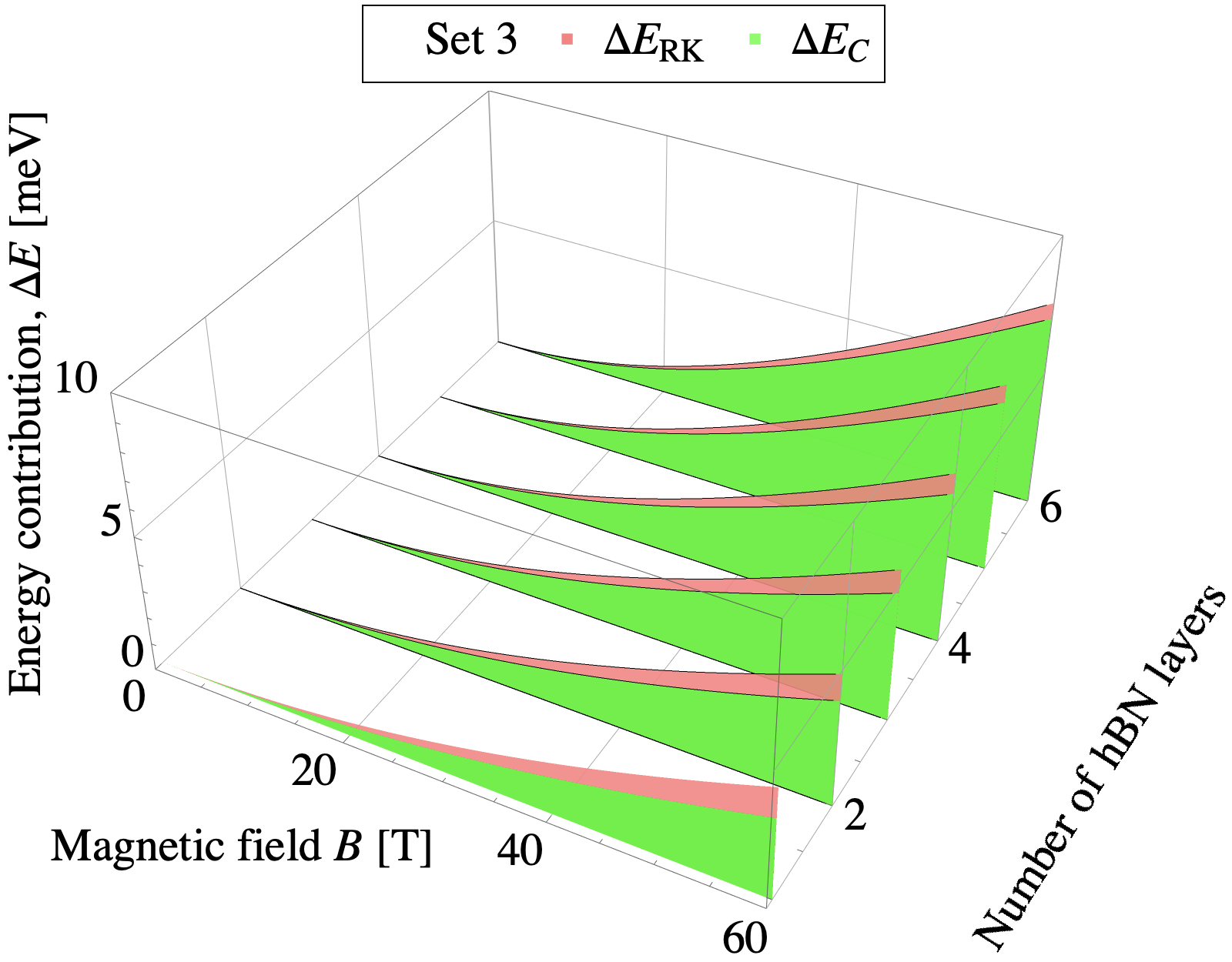} & %
\includegraphics[width=80mm]{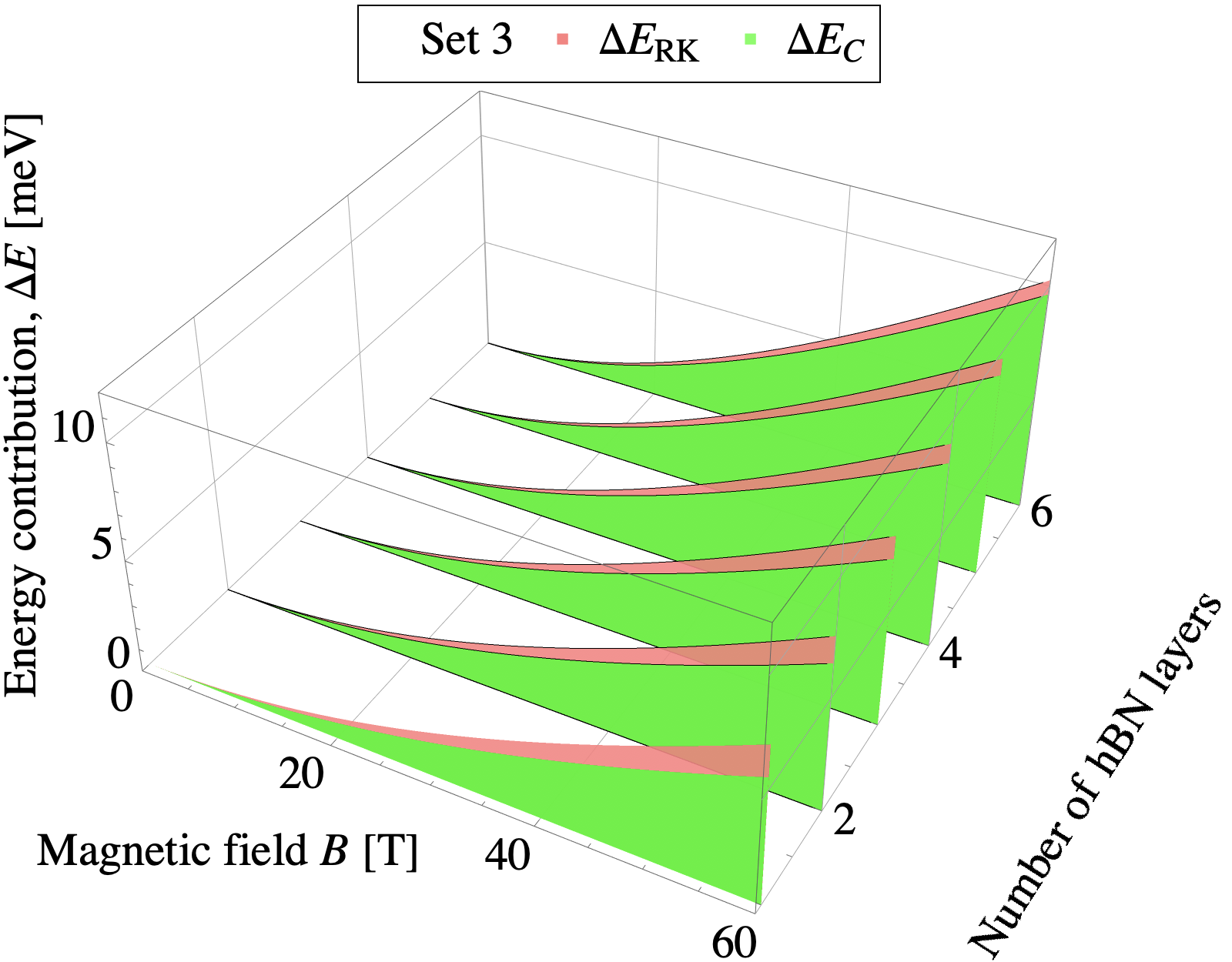} \\[6pt]
\textit{(c)} 3$s$ & \textit{(d)} 4$s$ \\[6pt]
\includegraphics[width=80mm]{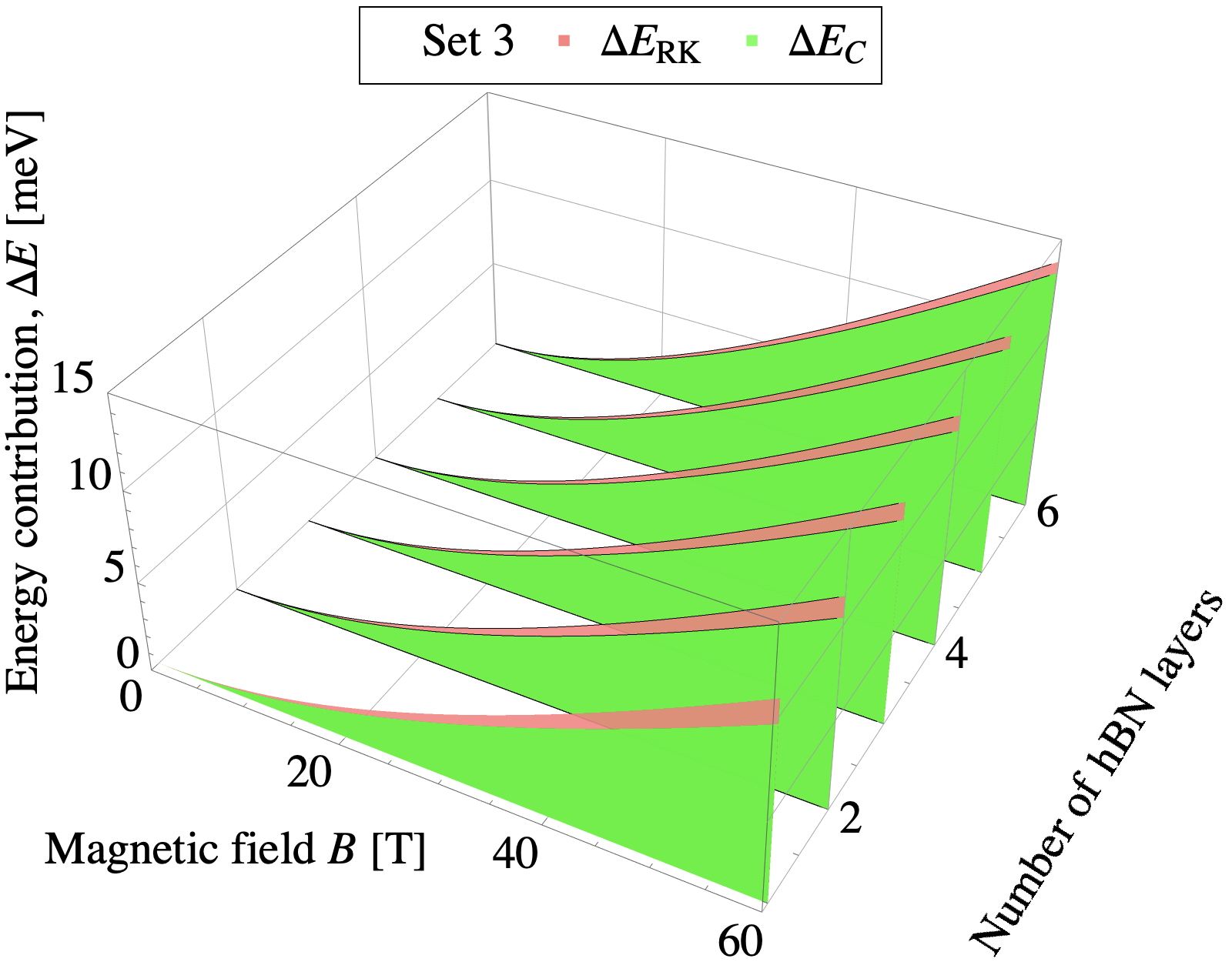} & %
\includegraphics[width=80mm]{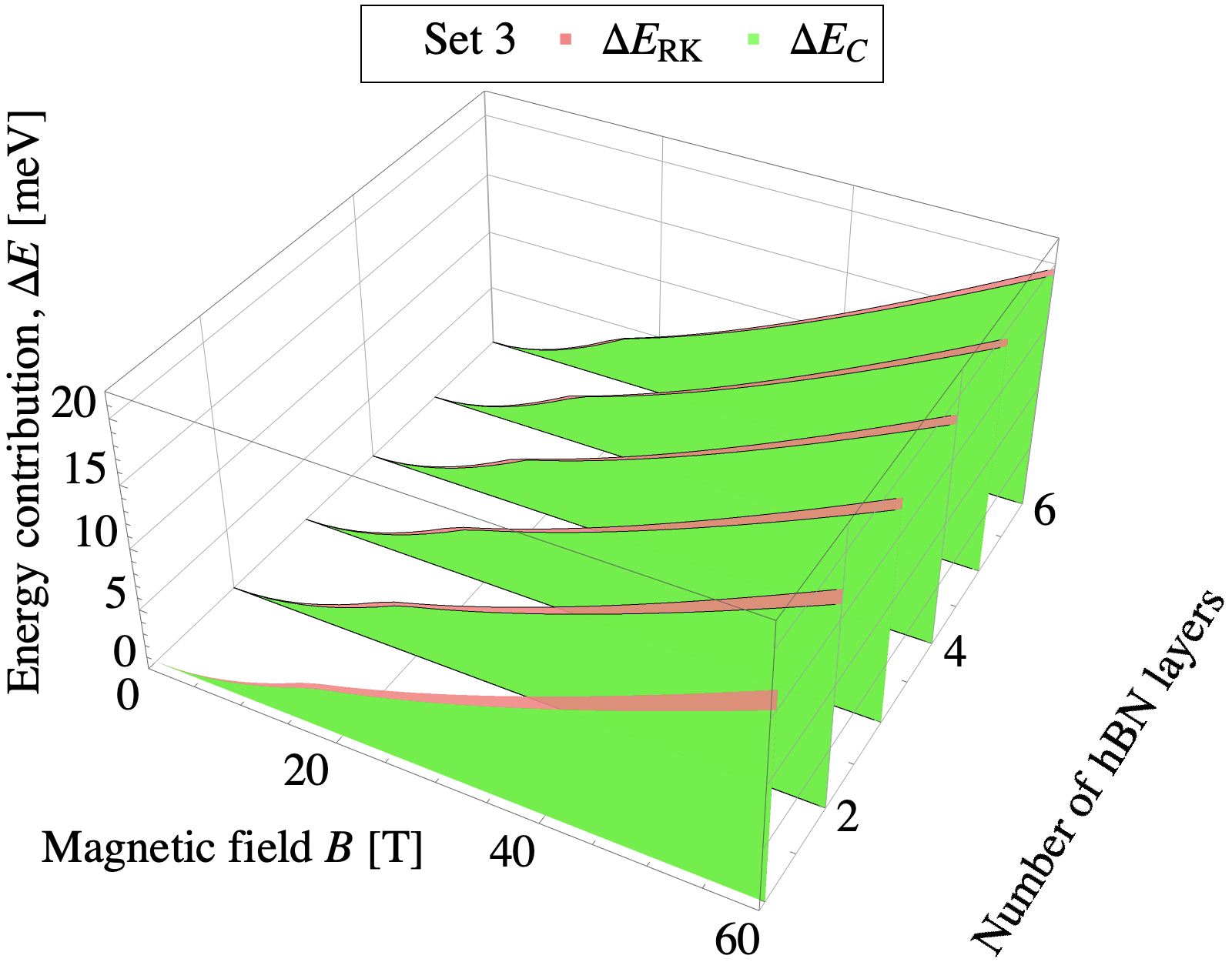} \\
&
\end{tabular}%
\caption{The energy contribution for indirect magnetoexcitons in the vdW heterostructures as a function of the magnetic field and the number of
hBN layers. The data are plotted for the set of masses 3. }
\label{het_set3}
\end{figure}
For the vdW heterostructure, in Fig. \ref{het_set3} we compare $%
\Delta E_{RK}$ and $\Delta E_C$. Results are given for the set 3 that is taken
to be the representative case. We plot $\Delta E$ as a function of the
external magnetic field and the number of hBN layers. As can be seen from
Fig. \ref{het_set3} as the number of hBN layers increases $\Delta E_{RK}$ and
$\Delta E_C$ converge. Once again, in contrast to indirect magnetoexcitons
in TMDCs and Xenes double-layer heterostructure, magnetoexcitons in the vdW phosphorene heterostructure stay bound in states 3$s$ and 4$s$
while the magnetic field is varied between 0 and 60 T. This can be
explained by the fact that the binding energies of states 3$s$ and 4$s$ are
higher for indirect magnetoexcitons in phosphorene due to their effectively quasi-one-dimensional nature. The quasi-one-dimensionality of the magnetoexcitons in phosphorene is also demonstrated by the comparison of the contribution of the external magnetic field to binding energies of the vdW phosphorene heterostructure shown in Fig. \ref{het_set3}  and TMDC heterostructures: $\Delta E$ due to the magnetic field in the vdW phosphorene heterostructure is always higher than the one in TMDC heterostructures.
We report in Table \ref{table:energies_het_set3} the binding energies of
Rydberg states 1$s$, 2$s$, 3$s$, and 4$s$ for $N$=1,2,3,4,5,6 when Eq. \ref%
{eq:finsch} is solved with $V_{RK}$ and $V_C$. The binding energies calculated using the Coulomb potential
are always larger than the one obtained with the RK potential. The increase of the phosphorene interlayer distance in the vdW heterostructure leads to the deduction of electron-hole interaction and results in the decrease of the binding energy of indirect magnetoexcitons.
\begin{table}[t]
\caption{Binding energies for indirect excitons in the van der Waals
heterostructures calculated using $V_{RK}$ and $V_C$ potentials for the set of masses 3. Energies are given in meV.}
\label{table:energies_het_set3}
\begin{center}
\sisetup{table-format=2.4}
\par
\begin{tabular}{P{1cm}P{1cm}P{1cm}P{1cm}P{1cm}|P{1cm}P{1cm}P{1cm}P{1cm}}
\hline \hline
\multirow{2}{*}{$N$}&
\multicolumn{4}{c|}{ $V_{RK}$}&
\multicolumn{4}{c}{$V_C$}
   \\ \cline{2-9}
   & 1$s$ & 2$s$ & 3$s$ & 4$s$ & 1$s$ & 2$s$ & 3$s$ & 4$s$ \\ \cline{1-9}
1 & 98.50 & 55.13 & 37.45 & 28.11 & 128.38 & 66.00 & 43.02 & 30.91 \\
2 & 86.39 & 51.29 & 35.27 & 27.15 & 104.90 & 58.98 & 39.30 & 29.39 \\
3 & 76.91 & 47.76 & 33.31 & 26.15 & 89.28  & 53.38 & 36.34 & 27.96 \\
4 & 69.32 & 44.60 & 31.55 & 25.16 & 78.03  & 48.84 & 33.90 & 26.64 \\
5 & 60.10 & 41.80 & 29.96 & 24.30 & 69.49  & 45.06 & 31.82 & 25.43  \\
6 & 57.93 & 39.31 & 28.53 & 23.29 & 62.75  & 41.88 & 30.03 & 24.32
\\   \hline \hline
\end{tabular}
\end{center}
\end{table}

%%%%%%%%%%%%%%%%%%%%%%%%%%%%%%%%%%%%%%%%%%%%%%%%%%%%%%%%%%%%%%%%%%%%%%%%%%%%%%%%%%%%%%%%%%%%%%%%%%%%%%%%%%%%%%%%%%%%%%%%%%%%%%%%%%%%%%%%%%%%%%%%%%%%%%%%%%%

%%%%%%%%%%%%%%%%%%%%%%%%%%%%%%%%%%%%%%%%%%%%%%%%%%%%%%%%%%%%%%%%%%%%%%%%%%%%%%%%%%%%%%%%%%%%%%%%%%%%%%%%%%%%%%%%%%%%%%%%%%%%%%%%%%%%%%%%%%%%%%%%%%%%%%%%%%%

\section{Diamagnetic coefficients} \label{coefficients}

In this paper, we calculate the diamagnetic coefficients in the same fashion as was done in
Refs. \cite{Spiridonova, Kez2021,RKAS2021b}. For phosphorene the diamagnetic coefficients for magnetoexcitons
Rydberg states are reported for the first time. The magnetic field range between 0 and
30 T has been used to calculated the DMCs. In this range of the magnetic field one can observe the linear dependence of energy on $B^{2}$. Here, we follow the notation in literature and denote
the diamagnetic coefficient as $\sigma$. The diamagnetic
coefficients for magnetoexcitons in monolayer and bilayer systems are
reported in Table \ref{table:diam_mon_b}. $\sigma$ for the indirect
magnetoexcitons in the vdW heterostructure are reported in Table %
\ref{table:diam_het}.
\begin{table}[h!]
\caption{The diamagnetic coefficients, $\protect\sigma$, of direct
magnetoexcitons in FS and encapsulated by hBN monolayers for the sets of masses 1 and 3. $%
\protect\sigma$ is given in $\protect\mu$eV/B$^2$. DMCs are obtained for the
range of the magnetic field between 0 and 30 T and correspond to $R^2$ =
0.9998 for the linear regression model. }
\label{table:diam_mon_b}
\begin{center}
\sisetup{table-format=2.4}
\par
\begin{tabular}{P{1cm}P{1.5cm}P{1.5cm}P{1.5cm}P{1.5cm}|P{1.5cm}P{1.5cm}P{1.5cm}P{1.5cm}}
\hline \hline
\multirow{3}{*}{State}&
\multicolumn{4}{c|}{Monolayer}&
\multicolumn{4}{c}{Bilayer} \\ \cline{2-9}
&\multicolumn{2}{c}{Set 1}&
\multicolumn{2}{c|}{Set 3} &
\multicolumn{2}{c}{Set 1} &
\multicolumn{2}{c}{Set 3}
   \\ \cline{2-9}
  &  hBN & FS & hBN & FS &  $V_{RK}$ & $V_C$ &  $V_{RK}$ & $V_C$ \\ \cline{1-9}
1$s$ & 0.94 & 0.41 & 0.46 & 0.21 & 0.97 & 0.44 & 0.52  & 0.23\\
2$s$ & 2.31 & 0.68 & 1.41 & 0.40 & 1.34 & 0.55 & 0.80  & 0.32\\
3$s$ & 4.13 & 0.99 & 2.96 & 0.67 & 1.81 & 0.71 & 1.22  & 0.47\\
4$s$ &      & 1.34 &      & 1.88 & 2.32 & 0.93 & 3.30  & 1.30
\\   \hline \hline

\end{tabular}
\end{center}
\end{table}

\begin{table}[t]
\caption{The diamagnetic coefficients, $\protect\sigma$, of indirect
magnetoexcitons in the vdW heterostructure for the sets of masses 1
and 3. $\protect\sigma$ is given in $\protect\mu$eV/B$^2$. DMCs are obtained
for the range of the magnetic field between 0 and 30 T and correspond to $%
R^2 $ = 0.9998 for the linear regression model.}
\label{table:diam_het}
\begin{center}
\sisetup{table-format=2.4}
\par
\begin{tabular}{P{1cm}P{1cm}P{1.5cm}P{1.5cm}P{1.5cm}P{1.5cm}}
\hline \hline
\multirow{2}{*}{State}&
\multirow{2}{*}{$N$}&
\multicolumn{2}{c}{Set 1}&
\multicolumn{2}{c}{Set 3}
   \\ \cline{3-6}
& &
\multicolumn{1}{c}{$V_{RK}$} &
\multicolumn{1}{c}{$V_{C}$} &
\multicolumn{1}{c}{$V_{RK}$} &
\multicolumn{1}{c}{$V_{C}$}
   \\ \cline{1-6}

\multirow{6}{*}{1$s$} & 1 & 2.36 & 1.72 & 1.24 & 0.88 \\
& 2 & 2.80 & 2.22 & 1.49 & 1.16 \\
& 3 & 3.23 & 2.72 & 1.74 & 1.45 \\
& 4 & 3.67 & 3.20 & 2.00 & 1.73  \\
& 5 & 4.09 & 3.68 & 2.27 & 2.02  \\
& 6 & 4.51 & 4.14 & 2.53 & 2.30 \\ \cline{1-6}
\multirow{6}{*}{2$s$} & 1 & 3.64 & 2.85 & 2.24 & 1.73\\
& 2 & 4.04 & 3.37 & 2.50 & 2.06 \\
& 3 & 4.46 & 3.88 & 2.77 & 2.39 \\
& 4 & 4.88 & 4.38 & 3.05 & 2.71 \\
& 5 &      &      & 3.34 & 3.03 \\
& 6 &      &      & 3.62 & 3.35\\ \cline{1-6}
3$s$ & 1 & & & & 3.39
\\   \hline \hline

\end{tabular}
\end{center}
\end{table}
We report the diamagnetic coefficients in Table \ref{table:diam_mon_b} for
two sets of masses. $\sigma$ for the FS and encapsulated by hBN
monolayers phosphorene is obtained when the magnetoexcitons interact through the $V_{RK}$ interaction, while for the bilayer $\sigma$ is given when Eq. (\ref%
{eq:finsch}) is solved using both $V_{RK}$ and $V_C$ potential. In the monolayer $%
\sigma_1>\sigma_3$, where $\sigma_1$ and $\sigma_3$ are DMCs obtained with sets of masses 1 and 3, respectively. For phosphorene encapsulated by hBN, $\sigma$
can be extracted in states 1$s$, 2$s$, and 3$s$. In the case of FS
phosphorene $\sigma$ can also be extracted in the state 4$s$. In the case of
bilayer phosphorene, the diamagnetic coefficients can be extracted for 1$s$, 2$s$, 3$s$,
and 4$s$ states. This is different to the diamagnetic coefficients of
bilayer composed of Xenes \cite{Kez2021} and TMDCs \cite{RKAS2021b}, where
$\sigma$ cannot be extracted for all examined states.

The DMCs for the vdW heterostructure are reported in
Table \ref{table:diam_het} for two sets of masses when Eq. (\ref%
{eq:finsch}) is solved using $V_{RK}$ and $V_C$ potentials. $\sigma$ can be
extracted for the state 1$s$ for all values of examined $N$, in the state 2$s$ for
the set 3 $\sigma$ can be extracted for $N=1,...,6$. But for the set 1 in the state 2$s$,
$\sigma$ can be extracted only for $N=1,...,4$. From data listed in Tables \ref{table:diam_mon_b} and \ref{table:diam_het}, it follows that $\sigma_{RK}>\sigma_C$, where $\sigma_{RK}$ and $\sigma_{C}$ are the DMCs obtained using the RK and Coulomb potentials, respectively. Finally, our calculations show that the magnetoexcitons binding energies and DMCs in
phosphorene vdW heterostructure can all be broadly tuned by changing the number of stacked hBN layers. This serves as a convenient and efficient method for engineering the materials properties.

%\nocite{*}

\section{Conclusions}
In the framework of the excitonic Mott-Wannier model, we study direct and indirect excitons in Rydberg states in phosphorene monolayers, bilayer, and vdW heterostructure in the external magnetic field, applied perpendicular to the monolayer or heterostructure. We calculated the binding energies and DMCs for the Rydberg states 1$s$, 2$s$, 3$s$, and 4$s$ for magnetoexcitons formed in these systems.

The magnetic field contribution to the binding energy of magnetoexcitons strongly depends on the effective masses of electron and hole and their anisotropy. Interestingly, the binding energies of magnetoexcitons are strongly correlated with the reduced mass anisotropy that makes magnetoexcitons in phosphorene effectively quasi-1D quasiparticles. An overall, the unique anisotropic character of these magnetoexcitons in phosphorene leads to the larger contribution to the binding energy than in TMDC materials.

The DMCs demonstrate the strong dependence on the effective electron-hole masses: DMCs in phosphorene monolayers, bilayer, and vdW heterostructure for the set 1 are about two fold bigger than the one for the set 2. The other distinct feature for the diamagnetic coefficients in the bilayer
and vdW heterostructures is that $\sigma$ calculated using the Coulomb potential is always smaller than DMCs obtained with the Rytova-Keldysh potential. In other word, the reduced dimensionality and screening increase the diamagnetic coefficients.

We show that the binding energy of direct and indirect magnetoexcitons can be tuned by the external magnetic field. Also, the binding energy of indirect excitons and DMC could be efficiently tuned by the stacking hBN layers. Such tunability of binding energies and DMCs is potentially useful for the devices design.

Finally, in summary, we have shown theoretically that the vdW phosphorene heterostructure is a novel category of 2D semiconductor offering a tunability of the binding energies of magnetoexcitons by mean of the external magnetic field and control by the number of hBN layers separating two phosphorene sheets. Also, the DMCs in the vdW heterostructure are tunable by controlling the number of hBN layers that separate two phosphorene layers. Thus, phosphorene provides a unique platform for novel optoelectronic applications and the exploration of the role of the
symmetry breaking in anisotropic exciton physics.

\

\textbf{Acknowledgments.} This work is supported by the U.S. Department of
Defense under Grant No. W911NF1810433 and PSC-CUNY Award No. 62261-00 50.
\bibliography{/Users/Nastia/Desktop/dissertation/bibliography/bibl_v3_phos}

\end{document}